\documentclass[conference,letterpaper]{IEEEtran}

\usepackage[utf8]{inputenc} 
\usepackage[T1]{fontenc}
\usepackage{url}
\usepackage{ifthen}
\usepackage{cite}
\usepackage[cmex10]{amsmath} 

\usepackage{amssymb}
\usepackage{graphicx}
\usepackage{cite}
\usepackage{physics}

\newtheorem{theorem}{Theorem}
\newtheorem{corollary}{Corollary}[theorem]
\newtheorem{proposition}[theorem]{Proposition}

\newtheorem{example}{Example}
\newtheorem{definition}{Definition}

\makeatletter
\NewCommandCopy\@@pmod\pmod
\DeclareRobustCommand{\pmod}{\@ifstar\@pmods\@@pmod}
\def\@pmods#1{\mkern4mu({\operator@font mod}\mkern 6mu#1)}
\makeatother

\begin{document}
\title{Higher-Order Staircase Codes: \\ A Unified Generalization of High-Throughput Coding Techniques} 

\author{
	\IEEEauthorblockN{Mohannad Shehadeh and Frank R.~Kschischang}
	\IEEEauthorblockA{University of Toronto\\
		The Edward S.\ Rogers Sr.\ Department of Electrical \& Computer Engineering\\
		Toronto, ON M5S 3G4, Canada\\
		Emails: \{mshehadeh, frank\}@ece.utoronto.ca}
}

\maketitle

\begin{abstract}
We introduce 
a unified generalization of several well-established 
high-throughput coding techniques including 
staircase codes, tiled diagonal zipper codes, 
continuously interleaved codes, open
forward error correction (OFEC) codes,
and Robinson--Bernstein convolutional
codes as special cases.
This generalization which we term ``higher-order staircase codes'' 
arises from the marriage 
of two distinct combinatorial objects:
difference triangle sets and finite-geometric nets, 
which have typically been applied separately 
to code design. We illustrate 
one possible realization of these codes, obtaining 
powerful, high-rate, low-error-floor, and low-complexity
coding schemes based on simple iterative syndrome-domain
decoding of coupled Hamming component codes. 
We study some properties of
difference triangle sets having minimum scope and sum-of-lengths, which correspond
to memory-optimal
higher-order staircase codes.
\end{abstract}

\section{Overview}\label{overview-section}
\IEEEPARstart{S}{taircase} codes \cite{smith} 
and related product-like codes 
\cite{sukmadji-cwit, OFEC, CI-patent, MC-TDZC}
provide a highly-competitive paradigm
for high-throughput error control coding.
Such codes use syndrome-domain iterative
bounded distance decoding of 
algebraic component
codes to achieve high decoder throughputs
while maintaining low internal decoder data-flow. 
Furthermore, their 
block-convolutional 
structures facilitate
pipelining and parallelism, the combination 
of which is essential for efficient hardware designs.
Moreover, their typically 
combinatorial constructions 
ensure control over error floors which 
must usually be below $10^{-15}$ 
in the relevant applications. 
Most notably, such codes are 
used in fiber-optic communications \cite[Ch.~7]{optical-networks-handbook}
but are also relevant to flash storage 
applications \cite{NAND-staircase} 
which increasingly demand 
similar code characteristics.

In this paper, we provide a
fruitful generalization of staircase codes 
which arises
from the combination of 
\emph{difference triangle sets} (DTSs) \cite[Part~VI:~Ch.~19]{Colbourn} 
and finite-geometric \emph{nets} 
\cite[Part~III:~Ch.~3]{Colbourn}.
Both DTSs and nets are well-studied 
combinatorial objects
and both have been separately
applied to code design, e.g., in \cite{CSOC,NB-DTS} and 
\cite{SJ-PG,TD-LDPC} respectively. In these
examples, DTSs lead naturally to convolutional
codes while nets lead naturally to block codes.
It is unsurprising then that their combination
leads to a generalization of staircase
codes which themselves combine aspects
of block and convolutional coding.

We provide a high-level summary of our construction
and contributions:
A \emph{higher-order staircase code} is determined
by three intertwined objects:  a DTS, a net,
and a \emph{component code}. Each of these
objects is characterized by two parameters
whose meanings are defined later.
The relationships between these parameters
and properties of the resulting code are
as follows:
Given an $(L,M)$-DTS, an $(M+1,S/L)$-net,
and an $((M+1)S, (M+1)S-r)$ component code, 
we get
a rate $1 - r/S$
spatially-coupled code 
in which every symbol
is protected by $M+1$ component
codewords and any pair of distinct 
component codewords share at most one symbol. 
Alternatively, 
the Tanner graph of the code 
is an infinite semiregular 
$4$-cycle-free bipartite graph 
with variable
node degree $M+1$ and generalized
constraint node degree $(M+1)S$.
The parameter $L$ controls the balance of
memory depth against parallelism 
while keeping all 
of the aforementioned properties completely
unchanged. A fourth parameter $C$
controls the parallelism obtained by circularly coupling multiple 
copies of a higher-order staircase code together. The parameters
$M$ and $C$ provide mechanisms for improving error floor 
and threshold performance respectively without altering
the component code. 

The proposed code family recovers
as particular cases some existing code families
and interpolates between them:
\begin{itemize}
	\item When $L = M = C = 1$, 
	the classical staircase codes of
	Smith et al.\ \cite{smith} are recovered.
	\item When $L = C = 1$ and $M \geq 1$, the generalized
	staircase codes of Shehadeh et al. \cite{mohannad-ofc} 
	are recovered.
	\item When $L \geq 1$ and $M = C = 1$, the
	tiled diagonal zipper codes 
	of Sukmadji et al.\ \cite{sukmadji-cwit} are recovered.
	\item When $L \geq 1$, $M = 1$, and $C = 2$, the open forward error correction (OFEC) code family \cite{OFEC} is recovered.
	\item When $L \geq 1$, $M = 1$, and $C \geq 2$, the multiply-chained tiled diagonal zipper codes of 
	Zhao et al.\ \cite{MC-TDZC} are recovered.
	\item When $S/L = M = C = 1$, the continuously interleaved codes of
	Coe \cite{CI-patent} are recovered.
	\item When $S/L = C = r = 1$, a
	recursive (rather than feedforward)
	version of Robinson--Bernstein convolutional codes \cite{CSOC} is recovered.
\end{itemize}
Importantly, when $M > 1$, we get a
higher-variable-degree generalization
of tiled diagonal zipper codes \cite{sukmadji-cwit}.
These codes are often described as providing an
encoding memory reduction by a factor of $1/2$ relative to
classical staircase codes
when $L \to \infty$ with $S/L$ held constant. 
We illustrate
analogous limits for higher-order staircase codes 
of $3/4$, $5/6$, and $9/11$ for
the cases of $M = 2$, $M = 3$, and $M = 4$ respectively.

\section{Code Construction}

Higher-order staircase codes can be interpreted
as convolutional codes in which coded bits
partition naturally into $S/L \times S/L$ blocks.
A codeword is then an infinite
sequence of $S/L \times S/L$ matrices $B_0, B_1, B_2, \dots$
satisfying certain constraints. For mathematical convenience,
we extend this sequence to be bi-infinite 
$\dots,B_{-2},B_{-1},B_{0},B_{1},B_{2},\dots$.
Our code is then defined by imposing constraints
on this sequence from which a recursive encoding
process can be derived.

We begin with the required objects. In what
follows, $[N]$ denotes the index set $\{0,1,\dots,N-1\}$.
A \emph{ruler} of \emph{order} $M+1$ 
is an ordered set of $M + 1$ \emph{distinct}
integers referred to as
\emph{marks}.
An \emph{$(L,M)$-DTS} is a set of $L$ rulers of order $M+1$
given by
$(d_0^{(\ell)}, d_1^{(\ell)}, \dots, d_M^{(\ell)})$
for 
$\ell \in [L]$  with the property that all differences $d^{(\ell)}_{k_1}-d^{(\ell)}_{k_2}$
for $k_1,k_2 \in [M+1]$
with $k_1 \neq k_2$ and $\ell \in [L]$ are \emph{distinct}.
A ruler is \emph{normalized} if 
$0 = d_0^{(\ell)} < d_1^{(\ell)} < \cdots < d_M^{(\ell)}$
and a DTS is normalized if all of its rulers are.
DTSs are typically constructed by computer search 
and we study some of
their relevant properties in Section \ref{DTS-section}.

Next, consider a set of $M+1$ permutations of $[S/L]\times [S/L]$ given by 
$\pi_k$ for $k \in [M+1]$. When an $S/L \times S/L$ matrix $B$ with 
variable entries $b_{(i,j)}$
for $i,j\in [S/L]$ is permuted according to $\pi_k$, the resulting matrix is denoted by $\Pi_k(B)$ and has entries given by $b_{\pi_k(i,j)}$ for $i,j\in [S/L]$. 
Such a set of permutations is
said to be an \emph{$(M+1,S/L)$-net} if, for any $k_1,k_2\in [M+1]$ with $k_1 \neq k_2$, any row of $\Pi_{k_1}(B)$ has only \emph{one} variable in common with any row of $\Pi_{k_2}(B)$. When $S/L = 1$, an $(M+1,1)$-net will comprise $M+1$ copies
of the identity permutation. Besides this case, these permutations
are necessarily distinct.

Nets are well-studied objects in combinatorics and finite
geometry with a variety of associated problems
as discussed in~\cite{Colbourn}. For our purposes,
it suffices to consider permutations
defined by linear transformations 
over a conveniently chosen finite commutative 
ring $\mathcal{R}$ of
cardinality $S/L$.
A permutation of $[S/L]\times [S/L]$
is interpreted as a permutation of 
$\mathcal{R} \times \mathcal{R}$
defined by left-multiplication of 
an invertible $2\times 2$ matrix over
$\mathcal{R}$.
These are trivial to characterize:

\begin{proposition}[Complete characterization of linear transformations
	defining $(M+1,S/L)$-nets]\label{intra-block-scattering-theorem}
	A collection of $M+1$ permutations of $\mathcal{R} \times \mathcal{R}$, 
	where $\mathcal{R}$ is a finite commutative ring of cardinality $S/L$, defined by
	left-multiplication of a set of invertible $2 \times 2$ matrices, 
	defines an $(M+1,S/L)$-net if and only if, for any pair of distinct matrices $A, \tilde A$ in the set
	where 
	\begin{equation*}
		A = 
		\begin{pmatrix}
			a & b \\
			c & d
		\end{pmatrix},\;
		\tilde A = 
		\begin{pmatrix}
			\tilde a & \tilde b \\
			\tilde c & \tilde d
		\end{pmatrix}\text{,}
	\end{equation*} 
	we have that $c\tilde d - d \tilde c$ is invertible 
	in $\mathcal{R}$.
\end{proposition}
\begin{IEEEproof}
	This follows by imposing that the 
	linear system $(i,j)A=(\tilde i,\tilde j)\tilde{A}$
	has a unique solution for $(j,\tilde j)$ 
	given fixed row indices $(i,\tilde i)$
	by rearranging and imposing an invertible determinant.
\end{IEEEproof}

In what follows, $\mathbb{Z}_{S/L}$ denotes the ring of integers
modulo $S/L$ while $\mathsf{lpf}(S/L)$ denotes the least prime
factor of $S/L$. By Proposition \ref{intra-block-scattering-theorem}
and elementary number theory, we get:

\begin{example}\label{net-perms}
	If $\mathcal{R}=\mathbb{Z}_{S/L}$ and $M \leq \mathsf{lpf}(S/L)$, then
	the identity matrix $I_{2\times 2}$
	together with 
	\begin{equation*}
		\begin{pmatrix}
			0 & 1\\
			1 & z
		\end{pmatrix}
	\end{equation*}
	for $z \in \{0,1,\dots, M-1\}$ define an $(M+1,S/L)$-net. 
\end{example}

\begin{example}\label{net-perms-invo}
	If $\mathcal{R}=\mathbb{Z}_{S/L}$ and $M \leq \mathsf{lpf}(S/L)$, then
	the identity matrix $I_{2\times 2}$
	together with the involutions
	\begin{equation*}
		\begin{pmatrix}
			-z & 1-z^2 \\
			1 & z
		\end{pmatrix}
	\end{equation*}
	for $z \in \{0,1,\dots, M-1\}$ define an $(M+1,S/L)$-net. 
\end{example}

Lastly, we require an \emph{$((M+1)S, (M+1)S-r)$ component code} 
$\mathcal{C}$ which is a linear, systematic code of 
length $(M+1)S$ and dimension $(M+1)S - r$, typically binary.
We can then provide our key definition:

\begin{definition}[Higher-order staircase code]\label{higher-order-staircase-code-def}
	Given all three of the following: 
	\begin{itemize}
		\item an $(L,M)$-DTS given by $
		0 = d^{(\ell)}_0 < d^{(\ell)}_1 < \dots < d^{(\ell)}_{M}
		$
		for $\ell\in [L]$ and a resulting ruler of order $L(M+1)$
		$d_0 < d_1 < \cdots < d_{L(M+1)-1}$
		obtained accordingly as
		\begin{multline*}
			\{d_k \mid k \in [L(M+1)]\} = \\
			\{Ld^{(\ell)}_k + \ell \mid k \in [M+1], \ell \in [L]\}\text{;}
		\end{multline*}
		
		\item an $(M+1,S/L)$-net with corresponding  
		$M+1$ permutations of $[S/L]\times [S/L]$ given by $\pi_k$
		for $k \in [M+1]$ where $\pi_0$ is the identity permutation,
		and a resulting collection of 
		$L(M+1)$ permutations of $[S/L]\times [S/L]$ 
		given by $\pi'_{k'} = \pi_{k}$ for 
		every $k\in [M+1]$ and 
		$k' \in [L(M+1)]$ such that 
		$d_{k'} \in \{Ld^{(\ell)}_k + \ell \mid \ell \in [L]\}$\text{; and}
		\item an $((M+1)S, (M+1)S-r)$ component code $\mathcal{C}$,
	\end{itemize}
	a \emph{higher-order staircase code} of 
	rate $1-r/S$ is defined
	by the constraint on the bi-infinite sequence of 
	$S/L \times S/L$ matrices 
	$\dots,B_{-2},B_{-1},B_{0},B_{1},B_{2},\dots$ that the rows of 
	\begin{equation*}
		\big(
		\Pi'_{L(M+1)-1}(B_{n-d_{L(M+1)-1}}) \big\vert
		\cdots \big\vert
		\Pi'_{1}(B_{n-d_1}) \big\vert
		\Pi'_{0}(B_{n-d_0})
		\big)
	\end{equation*}
	belong to $\mathcal{C}$ for all $n \in L\mathbb{Z}$.
\end{definition}

The significance of this definition is not only in the recovery of 
the codes listed in Section \ref{overview-section} as special cases,
but in that the use of proper DTSs and nets results in the following:
\begin{proposition}
	For a higher-order staircase code,
		distinct component code constraints share at most \emph{one} variable
		and 
		each variable (or coded symbol) is protected by $M+1$ component code constraints.
\end{proposition}

\begin{corollary}
	For a higher-order staircase code with a $t$-error-correcting component code,
	the Hamming weight of an uncorrectable 
	error under iterative
	bounded distance decoding of the component code constraints is at least $(M+1)t+1$.
\end{corollary}

In particular, these properties are inherited from classical
staircase codes \cite{smith} and generalized to higher variable degree $M + 1$.
We omit the proofs, but they are elementary.

\begin{figure}[t]
	\centering
	\includegraphics[width=0.5\columnwidth]{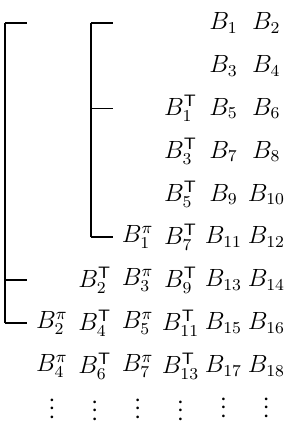}
	\caption{Visualization of the code of Example \ref{hosc-example}
		where $L = 2$ and $M = 2$;
		rows belong to $\mathcal{C}$ in this picture. }\label{hosc-example-fig}
\end{figure}

\begin{example}\label{hosc-example}
	Let $L = 2$ and $M = 2$. The
	$L = 2$
	rulers of order $M+1=3$ given by $
		(d_0^{(0)}, d_1^{(0)},d_2^{(0)}) = (0,6,7)$
		and 
		$
		(d_0^{(1)}, d_1^{(1)},d_2^{(1)}) = (0,2,5)$
	are a $(2,2)$-DTS.
	Consider the $(3,S/2)$-net
	of Example 
	\ref{net-perms} where $\pi_0(i,j)=(i,j)$, $\pi_1(i,j)=(j,i)$,
	and $\pi_2(i,j) = (j,i+j) \pmod{S/2}$ so that
	$\Pi_0(B) = B$ and $\Pi_1(B) = B^\mathsf{T}$. 
	We define $B^\pi = \Pi_2(B)$.
	A higher-order staircase code is defined by
	the constraint that the rows of 
	\begin{equation*}
		\big(B_{n-14}^\pi \big\vert B_{n-12}^\mathsf{T} \big\vert B_{n-11}^\pi 
		\big\vert B_{n-5}^\mathsf{T}\big\vert B_{n-1}\big\vert B_n\big)
	\end{equation*}
	belong to some component code $\mathcal{C}$
	for all $n \in 2\mathbb{Z}$. A recursive encoding process
	entails populating the first $S-r$ columns of $(B_{n-1}| B_n)$
	with information symbols, adjoining four previous blocks as
	above, and computing $r$ parity columns with a systematic
	encoder for $\mathcal{C}$, yielding an overall rate of $1-r/S$. 
	A visualization is provided
	in Fig.~\ref{hosc-example-fig}.
\end{example}

Next, we adapt a method of Zhao et al.\ \cite{MC-TDZC}
in which multiple tiled diagonal zipper codes 
\cite{sukmadji-cwit} are chained in a circle to
boost their performance. This straightforwardly extends
to higher-order staircase codes:
We constrain the $C$ bi-infinite sequences of 
	$S/L \times S/L$ matrices 
	$\dots,B_{-2}^{(c)},B_{-1}^{(c)},B_{0}^{(c)},B_{1}^{(c)},B_{2}^{(c)},\dots$ indexed by $c\in [C]$ so that the rows of
	\begin{multline*} 
	\bigg(
	\Pi'_{L(M+1)-1}(B_{n-d_{L(M+1)-1}}^{(c-1\pmod*{C})}) \bigg\vert\\ 
	\cdots \bigg\vert
	\Pi'_{L}(B_{n-d_{L}}^{(c-1\pmod*{C})}) \bigg\vert
	\Pi'_{L-1}(B_{n-d_{L-1}}^{(c)}) \bigg\vert
	\cdots
	\bigg\vert
	\Pi'_{0}(B_{n-d_0}^{(c)})
	\bigg)
	\end{multline*}
	for each $c\in [C]$ belong to $\mathcal{C}$ for all $n \in L\mathbb{Z}$.
Moreover, rather than considering a fixed component code $\mathcal{C}$, 
we can allow this code to vary as a function of row, $c$, and $n$ in the above.
This allows us to subsume the OFEC code family \cite{OFEC} in which the component
code is permuted during the encoding process.

Under an encoder style considered in \cite{sukmadji-cwit} 
analogous to Forney's \emph{obvious realization} 
\cite{Forney-Convolutional} of a convolutional encoder, the encoding
memory is proportional to $\sum_{\ell=0}^{L-1} d_M^{(\ell)}$. On the other hand,
the decoding memory, as well as the encoding memory
under other encoder styles, is 
proportional to $\max_{\ell\in [L]} d_M^{(\ell)}$. 
A \emph{memory-optimal} higher-order staircase code 
is one for which the DTS minimizes these two quantities.
If a tradeoff exists between the two, 
we take this to mean Pareto optimality.

\section{Simulations}\label{hosc-sim-section}

\begin{figure}[t]
	\centering
	\includegraphics[width=\columnwidth]{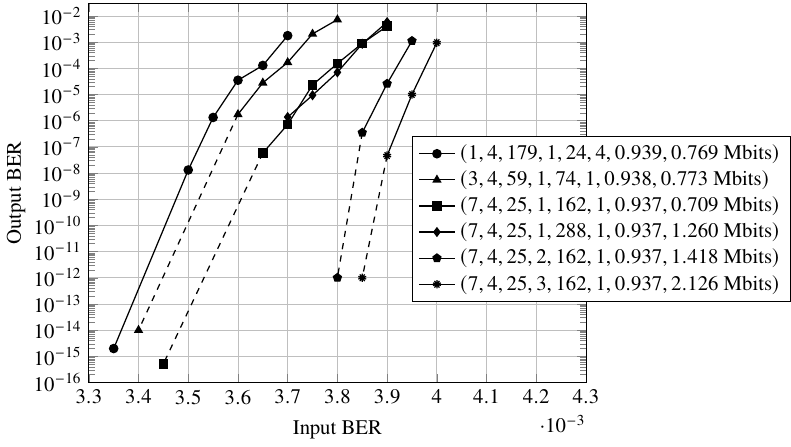}
	\caption{Simulation results for $1-r/S \approx 0.937$ examples with parameters $(L,M,S/L,C,W,I,1-r/S,WC(S/L)^2L)$.}\label{094-sims}
\end{figure}

\begin{figure}[t]
	\centering
	\includegraphics[width=\columnwidth]{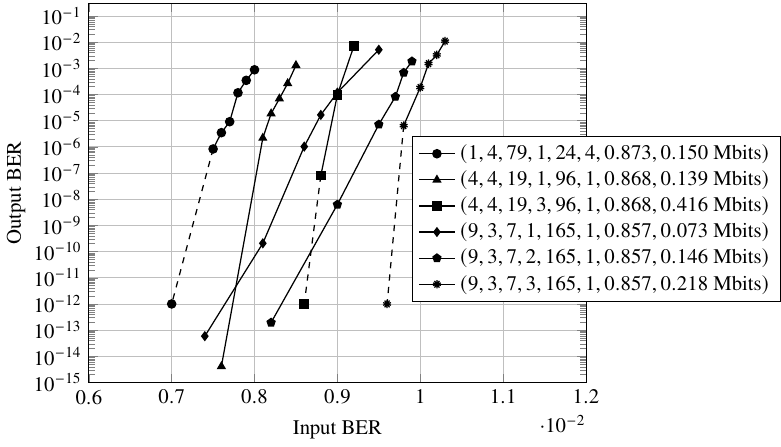}
	\caption{Simulation results for $1-r/S \approx 0.87$ examples with parameters $(L,M,S/L,C,W,I,1-r/S,WC(S/L)^2L)$.}\label{087-sims}
\end{figure}

We consider the use of memory-optimal 
higher-order staircase codes
with extended Hamming component codes
and binary symmetric channel transmission.
DTSs are found with a computer search
technique omitted from this paper.
The encoder produces $C(S/L) \times L (S/L)$ 
rectangles of coded bits 
and the parameter $W$ counts the number of such 
rectangles in the decoding window. 
We
perform sliding window decoding of 
$W$ consecutive rectangles where one iteration 
comprises decoding all constraints in the window consecutively 
and $I$ iterations are performed
before advancing the window by one rectangle. A 
termination-like procedure is used 
as in \cite{mohannad-ofc}. 
This yields a six-tuple of parameters $(L,M,S/L,C,W,I)$ with
which we label the curves
in Fig.~\ref{094-sims} and Fig.~\ref{087-sims}.
We also append the resulting code rate 
and decoding window size in bits.

The dashed lines indicate points at which
zero bit errors were measured after transmitting a number
of bits equal to ten times the reciprocal of the corresponding output bit error rate 
(BER). Fig.~\ref{087-sims} illustrates control over the error floor
and threshold via $M$ and $C$ respectively. 
Moreover, 
the codes in Fig.~\ref{094-sims} and Fig.~\ref{087-sims} perform
comparably to the codes considered in \cite{smith,MC-TDZC} 
with lower memory and lower complexity due to the 
use of Hamming components as opposed to BCH components. 
These results are obtained with highly-optimized,
high-throughput simulators which we provide
\cite{decsim-code,hosc-software}.

\section{Difference Triangle Sets with Minimum Scope and Sum-of-Lengths}\label{DTS-section}

A well-studied
hard problem is that
of constructing an $(L,M)$-DTS 
of minimum \emph{scope} \cite{Colbourn-IT,Klove}
where the \emph{scope} of a normalized DTS is its largest
mark.
Equivalently, scope is the largest 
\emph{length} among all rulers where the 
\emph{length} of a ruler is its largest difference which is
$\max_{\ell\in [L]} d_M^{(\ell)}$ for a normalized DTS. 
We have a \emph{trivial bound} 
on the scope of a normalized $(L,M)$-DTS:
\begin{equation}\label{scope-lower-bound}
	\max_{\ell\in [L]} d_M^{(\ell)}
	\geq 
	L \cdot \frac{(M+1)M}{2}
\end{equation}
with equality if and only if 
the set of positive differences between
all mark pairs, henceforth referred to as
the \emph{set of distances},  
is precisely $\{1,2,\dots,L(M+1)M/2\}$.
A DTS for which this bound 
holds with equality is said to be \emph{perfect}.

On the other hand, we also require
a minimization of the 
\emph{sum-of-lengths} 
$\sum_{\ell=0}^{L-1} d_M^{(\ell)}$.
To the
best of the authors' knowledge, this
problem has not been explicitly studied.
It is also \emph{not} equivalent to the minimum 
scope problem in general.

Except for the cases of $M = 1$
and $M = 2$, it remains an open problem to 
exhibit minimum scope $(L,M)$-DTSs for arbitrary 
$L$. However, perfect
DTSs admit recursive constructions which lead to
constructive existence results for 
infinitely many values of $L$. 
It is known that perfect DTSs
can only exist for $M \leq 4$. 
Therefore, for the cases of 
$M = 3$ and $M = 4$, recursive
constructions provide
the strongest existence results. 
We will study the implications
of one such construction due to 
Wild \cite{Wild} and Kotzig and Turgeon \cite{Kotzig-Turgeon}
for the problem of minimizing sum-of-lengths. We will
need the following:

\begin{definition}[Sharply $2$-transitive permutation group]
	A \emph{sharply $2$-transitive permutation group
		$G$ acting on a set $X$} 
	is a \emph{group}
	of permutations of $X$
	with the following property: 
	For each pair of ordered pairs 
	$(w,x), (y,z) \in X\times X$ with $w\neq x$
	and $y\neq z$, there exists a \emph{unique}
	permutation $g\in G$ such that
	$(g(w),g(x)) = (y,z)$.
\end{definition}

As noted
in \cite{Wild}, we can exhibit a
finite such group
as the group of invertible
affine transformations from a finite
field $\mathbb{F}_{M+1}$ to itself $x \mapsto ax+b$
where $M+1$ is a prime power and 
$a,b \in \mathbb{F}_{M+1}$ with $a\neq 0$
yielding $M(M+1)$ such permutations. 

\begin{theorem}[Special case of the combining construction of
	Wild \cite{Wild} and Kotzig and Turgeon \cite{Kotzig-Turgeon}]\label{combining-construction}
	Let $M+1$ be a prime power and let the collection of $M(M+1)$ permutations
	of $[M+1]$ given by $\rho_k$
	where $k\in [M(M+1)]$ be 
	a sharply $2$-transitive permutation group acting on $[M+1]$.
	Let $\mathcal{X}$ be a perfect $(L_1,M)$-DTS and 
	$\mathcal{Y}$ 
	be a perfect $(L_2,M)$-DTS with each given respectively by 
	$
		\mathcal{X} = \{(x_0^{(i)},x_1^{(i)},\dots,x_M^{(i)}) \mid i \in [L_1] \}
	$
	and
	$
		\mathcal{Y} = \{(y_0^{(i)},y_1^{(i)},\dots,y_M^{(i)}) \mid i \in [L_2] \}\text{.}
	$
	Let $\mathcal{W}$ be the set of $L_1L_2M(M+1)$ rulers given by
	\begin{multline*}
		(L_2M(M+1)+1)\left(x_0^{(i)},x_1^{(i)},\dots,x_M^{(i)}\right)
		+\\
		\left(y_{\rho_k(0)}^{(j)},y_{\rho_k(1)}^{(j)},\dots,y_{\rho_k(M)}^{(j)}\right)
	\end{multline*}
	for $i\in [L_1]$, $j\in [L_2]$, and $k\in [M(M+1)]$.
	Let $\mathcal{Z}$ be the union of $\mathcal{Y}$, $(L_2M(M+1)+1)\mathcal{X}$, and $\mathcal{W}$.
	Then, $\mathcal{Z}$ is a perfect $(L_1L_2M(M+1) + L_1 + L_2, M)$-DTS.
\end{theorem}

\begin{proposition}[Sum-of-lengths under combining]\label{sum-of-lengths-under-combining}
	Combining a perfect $(L_1,M)$-DTS with sum-of-lengths $S_1$ and a perfect 
	$(L_2,M)$-DTS with sum-of-lengths $S_2$ as in Theorem \ref{combining-construction} yields a perfect $(L_1L_2M(M+1)+L_1+L_2, M)$-DTS with sum-of-lengths $(L_2M(M+1)+1)^2S_1+S_2$.
\end{proposition}

\begin{IEEEproof}
	Trivially, $\mathcal{Y}$ and $(L_2M(M+1)+1)\mathcal{X}$ contribute
	$S_2$ and $(L_2M(M+1)+1)S_1$ to the sum-of-lengths so it remains
	to calculate the contribution of $\mathcal{W}$. The difference
	between the $v$th and $u$th mark of the ruler in
	$\mathcal{W}$ corresponding to a fixed
	$i\in [L_1]$, $j\in [L_2]$, and $k\in [M(M+1)]$
	can be expressed as 
	\begin{equation*}
		(L_2M(M+1)+1)\left(x_v^{(i)}-x_u^{(i)}\right) + \left(y_{\rho_k(v)}^{(j)}-y_{\rho_k(u)}^{(j)}\right)\text{.}
	\end{equation*}
	Since the term $y_{\rho_k(v)}^{(j)}-y_{\rho_k(u)}^{(j)}$ satisfies
	\begin{equation*}
		-\frac{L_2M(M+1)}{2} \leq y_{\rho_k(v)}^{(j)}-y_{\rho_k(u)}^{(j)} \leq \frac{L_2M(M+1)}{2}\text{,}
	\end{equation*}
	the scale factor $(L_2M(M+1)+1)$ guarantees that the ordering
	of the differences $x_v^{(i)}-x_u^{(i)}$ is preserved. Assuming
	without loss of generality that the rulers of $\mathcal{X}$
	are normalized, the 
	lengths of the corresponding rulers in $\mathcal{W}$
	are given by
	\begin{equation}\label{ruler-length-in-W}
		(L_2M(M+1)+1)x_M^{(i)} + \left(y_{\rho_k(M)}^{(j)}-y_{\rho_k(0)}^{(j)}\right)\text{.}
	\end{equation}
	By sharp $2$-transitivity, we have 
	$$
		\sum_{k = 0}^{M(M+1)-1}
		y_{\rho_k(M)}^{(j)}-y_{\rho_k(0)}^{(j)} = 0
	$$
	in which case 
	summing \eqref{ruler-length-in-W}
	over all $i\in [L_1]$, $j\in [L_2]$, and $k\in [M(M+1)]$,
	yields 
	$
		L_2M(M+1)(L_2M(M+1)+1)S_1
	$
	which is then added to $S_2$ and $(L_2M(M+1)+1)S_1$.
\end{IEEEproof}

We will now consider each of the cases of
$M\in\{1,2,3,4\}$ which are both theoretically
interesting for the reasons just discussed, 
but also most relevant to our proposed codes.
We will assume normalized DTSs henceforth.

For $(L,1)$-DTSs, 
we have the trivial bound $\max_{\ell\in [L]} d_1^{(\ell)} \geq L$
and trivially $\sum_{\ell=0}^{L-1} d_1^{(\ell)} \geq L(L+1)/2$. Both of these bounds hold with equality
for the perfect $(L,1)$-DTS given by 
$\{(0,1),(0,2),\dots,(0,L)\}$.

For $(L,2)$-DTSs, it is known that parity considerations
slightly strengthen the trivial bound to
\begin{equation}\label{deg-3-scope-lb}
	\max_{\ell\in [L]} d_2^{(\ell)} \geq 
	\begin{cases}
		3L & \text{if } L \equiv 0 \text{ or } 1 \pmod*{4} \\
		3L + 1 & \text{if } L \equiv 2 \text{ or } 3\pmod*{4}
	\end{cases}\text{.}
\end{equation}
From 
\begin{equation*}
	2\sum_{\ell = 0}^{L-1} d_2^{(\ell)} = 
	\sum_{\ell = 0}^{L-1}
	((d_2^{(\ell)}-d_1^{(\ell)}) 
	+ d_1^{(\ell)}
	+ d_2^{(\ell)})
	\geq \frac{3L(3L+1)}{2}\text{,}
\end{equation*}
it follows that
\begin{multline*}
	\label{deg-3-slen-lb}
	\sum_{\ell=0}^{L-1} d_2^{(\ell)} \geq  
	\begin{cases}
		\frac{3L(3L+1)}{4} & \text{if } L \equiv 0 \text{ or } 1 \pmod*{4} \\
		\frac{(3L-1)3L}{4} + \frac{3L+1}{2} & \text{if } L \equiv 2 \text{ or } 3\pmod*{4}
	\end{cases}
\end{multline*}
where the first inequality holds with equality if
and only if the set of distances is precisely $\{1,2,\dots,3L\}$
and the second inequality holds with equality if and only if
the set of distances is precisely $\{1,2,\dots,3L-1,3L+1\}$.
Both sets of bounds are achieved simultaneously 
by the explicit
constructions of Skolem \cite{Skolem} and 
O'Keefe \cite{OKeefe}. However, 
note that the DTS of Example
\ref{hosc-example} shows that it is possible
to achieve one of these bounds without
achieving the other.

For $(L,3)$-DTSs, the trivial bound is
\begin{equation}\label{deg-4-scope-lb}
	\max_{\ell\in [L]} d_3^{(\ell)} \geq 6L
\end{equation}
and it is a longstanding conjecture 
of Bermond \cite{Bermond} that this is achievable
for all $L$ excluding $L\in \{2,3\}$. 
Theorem \ref{combining-construction}
implies that there are infinitely
many values of $L$ for which 
\eqref{deg-4-scope-lb} is achieved, by, 
for example, repeatedly
combining the perfect $(1,3)$-DTS
$\{(0,1,4,6)\}$ with itself.
Bermond's conjecture has been verified
for all $L \leq 1000$ in \cite{deg-4-1000}
by using a wide variety of recursive constructions
to cover gaps as needed. 

\begin{proposition}
	For an $(L,3)$-DTS, we have
	\begin{equation}\label{deg-4-slen-lb}
		\sum_{\ell = 0}^{L-1}d_3^{(\ell)} \geq 5L^2+L
		\text{.}
	\end{equation}
	\begin{IEEEproof}
		By the distinctness of positive differences, we have
		\begin{align*}
			\sum_{u=1}^{6L} u
			&\leq\sum_{\ell = 0}^{L-1}\sum_{i = 0}^{3}\sum_{j=i+1}^{3}
			d_j^{(\ell)}-d_i^{(\ell)}\\
			&= 
			4\sum_{\ell = 0}^{L-1} d_3^{(\ell)}-\sum_{\ell = 0}^{L-1}(d^{(\ell)}_1-d^{(\ell)}_0)+
			(d^{(\ell)}_3-d^{(\ell)}_2)\\
			&\leq 
			4\sum_{\ell = 0}^{L-1} d_3^{(\ell)}-\sum_{u=1}^{2L}u
		\end{align*}
		which yields the desired result after rearranging.
	\end{IEEEproof}
\end{proposition}

It is not the case that $(L,3)$-DTSs achieving 
\eqref{deg-4-scope-lb} such as those listed in 
\cite{Shearer-IBM-Optimal-List} achieve \eqref{deg-4-slen-lb}.
However, we conjecture that it is
possible to simultaneously achieve \eqref{deg-4-scope-lb}
and \eqref{deg-4-slen-lb} for all $L$ excluding
$L\in \{2,3,4,5\}$ from computer search evidence.

Next, we consider constructing
an infinite family via Theorem
\ref{combining-construction}.
For all integers $i \geq 0$, 
let $\mathcal{Z}_i$ be a 
perfect $(L_i,3)$-DTS
with sum-of-lengths 
$S_i$ obtained by combining 
$\mathcal{Z}_{i-1}$ and $\mathcal{Z}_{0}$
via Theorem \ref{combining-construction}
where $\mathcal{Z}_0$ is a fixed 
perfect $(L_0,3)$-DTS with sum-of-lengths
$S_0$. Proposition \ref{sum-of-lengths-under-combining}
provides linear recurrences for $L_i$ and $S_i$
which we can solve to get
$L_i = ((12L_0+1)^{i+1}-1)/12$
and $
	S_i = 
	S_0(6L_i^2
	+ 
	L_i)/(6L_0^2+L_0)$
If $\mathcal{Z}_0$ satisfies
$\eqref{deg-4-slen-lb}$ so that
$S_0 = 5L_0^2+L_0$, we get
\begin{equation*}
	S_i = 
	\left(5+\frac{1}{6L_0+1}\right)L_i^2
	+\left(1-\frac{L_0}{6L_0+1}\right)L_i\text{.}
\end{equation*}
The largest $L_0$ for which we found
a perfect $(L_0,3)$-DTS achieving \eqref{deg-4-slen-lb}
with a computer search technique is $L_0 = 15$. 
The best result we can get is therefore
as follows:
\begin{proposition}\label{infinite-deg-4-fam}
	There are infinitely many values of $L$ for which a
	perfect $(L,3)$-DTS with sum-of-lengths
	\begin{equation*}
		\left(5+\frac{1}{91}\right)L^2+\left(1-\frac{15}{91}\right)L
	\end{equation*}
	exists.
\end{proposition}

For $(L,4)$-DTSs, it is known that the
trivial bound can be slightly strengthened
by parity considerations to
\begin{equation}\label{deg-5-scope-lb}
	\max_{\ell\in [L]} d_4^{(\ell)} \geq 
	\begin{cases}
		10L & \text{if } L \equiv 0 \pmod*{2} \\
		10L + 1 & \text{if } L \equiv 1 \pmod*{2}
	\end{cases}\text{.}
\end{equation}

\begin{proposition}
	For an $(L,4)$-DTS, we have
	\begin{equation}\label{deg-5-slen-lb}
		\sum_{\ell=0}^{L-1} d_4^{(\ell)}\\ \geq  
		\begin{cases}
			9L^2+\frac{3}{2}L & \text{if } L \equiv 0 \pmod*{2} \\
			9L^2+\frac{3}{2}L + \frac{1}{2} & \text{if } L \equiv 1\pmod*{2}
		\end{cases}\text{.}
	\end{equation}
\end{proposition}
\begin{IEEEproof}
	By the distinctness of positive differences,
	we have
	\begin{multline*}
		2\sum_{\ell = 0}^{L-1} d_4^{(\ell)}
		= 
		\sum_{\ell = 0}^{L-1}\bigg[ 
		(d^{(\ell)}_2-d^{(\ell)}_0) +
		(d^{(\ell)}_4-d^{(\ell)}_2)+
		(d^{(\ell)}_1-d^{(\ell)}_0) +\\
		(d^{(\ell)}_2-d^{(\ell)}_1) +
		(d^{(\ell)}_3-d^{(\ell)}_2) +
		(d^{(\ell)}_4-d^{(\ell)}_3) 
		\bigg]
		\geq 
		\sum_{u=1}^{6L} u
	\end{multline*}
	which yields the result after integrality
	considerations.
\end{IEEEproof}

We conjecture that \eqref{deg-5-scope-lb}
and \eqref{deg-5-slen-lb} are simultaneously achievable
for all $L$ excluding $L\in \{2,3,4\}$. The problem of achieving \eqref{deg-5-slen-lb} was considered implicitly by
Laufer \cite{Laufer} who sought structured perfect $(8,4)$-DTSs
to simplify a computer search.
Finally, using a perfect $(10,4)$-DTS achieving \eqref{deg-5-slen-lb} due to Khansefid et al.\  \cite{KTG,Shearer-IBM-Optimal-List,Shearer-IBM-Optimal-List-Credits}, we get the following with the same technique used for Proposition 
\ref{infinite-deg-4-fam}:
\begin{proposition}\label{infinite-deg-5-fam}
	There are infinitely many values of $L$ for which a
	perfect $(L,4)$-DTS with sum-of-lengths
	\begin{equation*}
		\left(9+\frac{6}{101}\right)L^2+\left(\frac{3}{2}-\frac{60}{101}\right)L
	\end{equation*}
	exists.
\end{proposition}

Consider dividing the encoding
memory $(S/L)^2\sum_{\ell=0}^{L-1} d_M^{(\ell)}$ 
by its value in the $L=1$ case
and taking the limit as $L\to \infty$
with $S/L$ held constant (so that $S \to \infty$).
Assuming that for each $M\in\{1,2,3,4\}$, there
are infinitely many values
of $L$ for which the respective sum-of-lengths
lower bounds are met,
the resulting limits are $1/2$, $3/4$, $5/6$, and $9/11$
respectively.
This premise is only conjectured to be
true in the cases of $M\in\{3,4\}$ but Propositions
\ref{infinite-deg-4-fam} and \ref{infinite-deg-5-fam}
allow us to prove limits of $5/6 + 1/546$ and 
$9/11 + 6/1111$ in those cases instead.

\section{Future Work}\label{Conclusion}

Several directions for future work are immediately
apparent.
On the mathematical side, we can attempt to strengthen Propositions \ref{infinite-deg-4-fam}
and \ref{infinite-deg-5-fam} via the many other
construction methods for perfect DTSs provided in 
\cite{deg-4-1000}. On the coding side, we can consider
more thorough simulation-based and theoretical 
performance analysis of higher-order staircase codes.

\IEEEtriggeratref{14}


\end{document}